\documentclass[reprint,aps,prd,twocolumn,showpacs,showkeys, 10pt, longbibliography, nolinenumbers, floatfix]{revtex4-1}

\usepackage{amsmath}
\usepackage{amssymb}
\usepackage{amsfonts}
\usepackage{graphicx}
\usepackage{float}
\usepackage{dcolumn}
\usepackage{multirow}
\usepackage{natbib}
\usepackage{tcolorbox}
\usepackage{parskip}
\usepackage[colorlinks = true,linkcolor = blue,urlcolor  = blue,citecolor = blue,anchorcolor = blue]{hyperref}
\usepackage{enumerate}

\usepackage{times}

\begin{document}


\title{Thermal Quarkyonic Matter and Its Implications for Neutron Star Structure}
\author{K. Folias and Ch.C. Moustakidis}



\affiliation{Department of Theoretical Physics, Aristotle University of Thessaloniki, 54124 Thessaloniki, Greece 
}
\begin{abstract}
The structure and basic properties of dense nuclear matter still remain one of the open problems of Physics. In particular, the composition of the matter that composes neutron stars is under theoretical and experimental investigation. Among the theories that have been proposed, apart from the classical one where the composition is dominated by hadrons, the existence or coexistence of deconfined  quark matter is a dominant guess. An approach towards this solution is the phenomenological view according to which the existence of quarkyonic  matter plays a dominant role in the construction of the equation of state (EOS). According to it the structure of the EOS is based on the existence of the quarkyonic particle  which is a hybrid state of a particle that combines properties of hadronic and quark matter with a corresponding representation in momentum space. In this paper we propose a phenomenological model for hot quarkyonic matter, borrowed from corresponding applications in hadronic models, where the interaction in the quarkyonic  matter  depends not only on the position but also on the momentum of the quarkyonic particles. This consideration, as we demonstrate, can have a remarkable  consequence on the shape of the EOS and thus on the properties of neutron stars, especially in those for which the effect of temperature is significant, offering a sufficiently flexible model.  Comparison with recent observational data can place constraints on the parameterization of the particular model and help improve its reliability.


\end{abstract}

\maketitle


\section{Introduction}
One of the fundamental problems of Nuclear Physics and Astrophysics remains the composition of dense nuclear matter as well as its basic properties both at zero and at finite temperature~\cite{Glendenning1996CompactSN,Shapiro:1983du,Haensel2007NeutronS1,schaffner-bielich_2020}. In particular, the  equation of state   of neutron star matter is the key quantity to study  these objects. In recent years  important astrophysical observations concerning not only isolated neutron stars but also binary systems, such as for example the merger of such a system with parallel emission of gravitational waves, can help decisively in the deeper knowledge of the interior of these objects \cite{Bernuzzi-2020,Metzger-2002,Burns-2020}. In this effort, a key problem that often arises is the inability of the EOSs to predict maximum masses for neutron stars that are compatible with recent observations (well above two solar masses) without simultaneously violating the sound-speed causality.

An interesting attempt in this direction is to consider a hybrid state of dense nuclear matter called quarkyonic matter~\cite{PhysRevC.110.025201,PhysRevC.107.065201,Cao:2022inx,Kovensky:2020xif,Quiros:1999jp,Das:2000ft,Koch_2023,Koch_2025,Jeong:2019lhv,PhysRevLett.122.122701,McLerran:2020rnw,article,Fukushima_2016,PhysRevC.108.045202,PhysRevLett.132.112701,McLerran-2024,Kojo-2024,Kojo-2025,Bluhm-2024,Lysenko-2025,Gao-2025a,Gao-2025b,Moss-2025,Pang-2024,Yamamoto-2023,Ciao-2020,Sen-2021a,Ivanytskyi-2025,Zhao-2020,Margueron-2021,Kojo-2021,Abramchuk-2021,Folias-2025,Gartlein-2025,Sen-2021,Xia-2023,Day-2025,Kalita-2025}. In this consideration, quarkyonic matter both quarks and nucleons appear as quasiparticles. According to the analysis of Refs.~\cite{PhysRevLett.122.122701,McLerran:2020rnw,article}, the basic assumption of quarkyonic matter  is that at large Fermi energy, the degrees of freedom inside the Fermi
sea may be treated as quarks, and confining forces remain
important only near the Fermi surface where nucleons
emerge through correlations between quarks. In quarkyonic matter, confinement at the Fermi surface
produces triplets with spin $1/2$ that we identify with
baryons that  occupy  a momentum shell of width $\Delta \simeq \Lambda_{QCD}$ (where $\Lambda_{QCD}$ represents  the confinement
scale)~ \cite{PhysRevLett.122.122701,Fukushima_2016,PhysRevC.108.045202,PhysRevLett.132.112701,McLerran-2024}. The width of the momentum shell $\Delta$ depends on the baryon density. 

Although in the literature there has been, in recent years, extensive theoretical study of quarkyonic matter (QM), all of these studies concern cold nuclear matter (the only exception is the study~\cite{Sen-2021}). In any case,  it would be of interest to extend the calculations to the case of hot nuclear matter, and this for two main reasons. The first is that, in this way, the theory of quarkyonic matter attains completeness, and the second pertains to the fact that investigations of neutron stars at birth (proto-neutron stars), neutron-star cooling, neutron-star mergers, as well as heavy-ion collision experiments, all necessitate a detailed understanding of nuclear matter at finite temperatures.
Thus, the main motivation of this work is  to extend the model of cold quarkyonic matter, as defined in previous studies, so as to also include thermal effects, in a manner that is as self-consistent as possible. Thus, following the approach of our previous work, we propose a quarkynonic model  where the interaction between  baryons, depends not only on the position but also on the momentum and has the advantage  that can be extended from zero to finite temperature~\cite{Prakash-1997,Bertsch-1988,Gale-1987,Li-2008,Moustakidis-2007,Moustakidis-2008,Moustakidis-2009a,Moustakidis-2009b}. In particular, we examine a QM case that includes neutrons alongside up and down quarks. We account for the interaction of neutrons as being dependent on their momentum, while quarks are treated as non-interacting and follow the statistics of a free Fermi gas. In this study, we focus primarily on the effect of temperature on the equation of state of quarkyonic matter and less on its specific parametrization, which will be the subject of a future investigation. We also examine how temperature influences the composition of QM across different baryonic densities and its impact on the corresponding equations of state. In addition, we analyze how these effects shape key properties of hot neutron stars, including their mass, radius, and tidal deformability.

The paper is structured as follows: Section II outlines the two theoretical frameworks employed to describe hot pure neutron matter and hot quarkyonic matter, both of which incorporate momentum-dependent interactions. Section III provides a concise overview of neutron-star structure, while Section IV presents and discusses the results of the present analysis. Finally, Section V offers concluding remarks.

\vspace{-0.2cm}

\section{The model}

Below, we outline the key aspects of the finite-temperature pure neutron matter model, as well as those of the finite-temperature quarkyonic matter framework. Additional details regarding the pure neutron matter model can be found in Refs.~\cite{Prakash-1997,Bertsch-1988,Gale-1987,Li-2008,Moustakidis-2007,Moustakidis-2008,Moustakidis-2009a,Moustakidis-2009b}. For the quarkyonic matter model, we attempt to generalize the zero-temperature framework to finite temperatures~\cite{PhysRevLett.122.122701,McLerran:2020rnw,article}. This extension naturally necessitates the adoption of specific assumptions, which are delineated and examined in the following text.

\subsection{Hot pure neutron matter }
The nuclear model  used in the present work is
designed to reproduce the results of microscopic calculations
of both nuclear and neutron-rich matter at zero temperature
and can be extended to finite temperature~\cite{Prakash-1997,Bertsch-1988,Gale-1987,Li-2008,Moustakidis-2007,Moustakidis-2008,Moustakidis-2009a,Moustakidis-2009b}. According to this model the energy
density of the pure neutron  matter is given by
the relation
\begin{equation}
{\cal E}_n(n_n,T)=\epsilon_{\rm kin}(n_n,T)+V_{\rm int}(n_n,T).
\label{def-energy}    
\end{equation}
The contribution  of the kinetic part is
\begin{equation}
\epsilon_{\rm kin}(n_n,T)=\frac{g_s}{(2\pi)^3}\int d^3k \sqrt{(\hbar c k)^2+ m_{n}^2c^4}f_n(n_n,k,T),\end{equation}
while the contribution of the interaction part reads 
 \begin{eqnarray}
V_{\rm int}(n_n,T)&=&\frac{1}{3}An_0\left(1-x_0   \right)u^2 +\frac{\frac{2}{3}Bn_0\left(1-x_3\right)u^{\sigma+1} }{1+\frac{2}{3}B'\left(1-x_3   \right)u^{\sigma-1}}\nonumber\\
&+& u\sum_{i=1,2}\frac{1}{5}\left[\frac{}{}6C_i-8Z_i \right]{\cal J}_i(n_n,T),
\label{eq5}
\end{eqnarray}
where $n_0$ is the nuclear saturation density ($n_0=0.16 \ $fm$^{-3}$),  $u =n_n/n_0$ and $f_n(n_n,k,T)$ which defined as 
\begin{equation}f_n(n_n,k,T)= \frac{1}{e^{(e_n(k,n_n,T)-\mu_n)/T}} \label{eq2}\end{equation}
is the Fermi-Dirac distribution function (hereinafter, for the sake of brevity, will simply be denoted as $f_n$). Moreover, $e_n(k,n_n,T)$ is the single particle energy and  $\mu_n$  stands for the  chemical potential of neutrons.
The function ${\cal J}_i(n_n,T)$ is given by 
\begin{eqnarray}
{\cal J}_i(n_n,T)&=&\frac{2}{(2\pi)^3}\int f_n {\rm g}(k,\Lambda_i)  d^3k, 
\label{Jn-1}
\end{eqnarray}
where the function ${\rm g}(k,\Lambda_i)$, suitable for simulating 
finite-range effects, is of the form~\cite{Prakash-1997,Moustakidis-2008}
\begin{equation}
{\rm g}(k,\Lambda_i)=\left[1+\left(\frac{k}{\Lambda_i}  \right)^2 \right]^{-1}.  
\label{g-1}
\end{equation}
The neutron number density is defined as  
\begin{equation} n_n = \frac{g_s}{(2\pi)^3}\int d^3k f_n(n_n,k,T),
\label{eq1}
\end{equation}
where $g_s$ is the degeneracy of the spin, that is $g_s=2$. In particular, the single particle energy is written as 
\begin{equation}
e_n(k,n_n,T)=e_n(k)+U(k,n_n,T), 
\label{SPE-1}
\end{equation}
where
$e_n(k)$ is the kinetic part of the single particle energy and  $U(k,n_n,T)$ is the sigle-particle potential.  For the kinetic part of the single particle energy we will use the relativistic expression
\begin{equation}e_n(k) = \sqrt{(\hbar k c)^2 + m_n^2 c^4},
\label{eq3} \end{equation}
where $m_n$ is the neutron mass. The single particle potential $U(k,n_n,T)$ is obtained from the functional derivative of the
interaction part of the energy density $V_{\rm int}(n_n,T)$ with respect to Fermi-Dirac distribution function $f_n$ and takes the form~\cite{Prakash-1997,Bertsch-1988,Gale-1987,Li-2008,Moustakidis-2007,Moustakidis-2008,Moustakidis-2009a,Moustakidis-2009b}
\begin{eqnarray}
U(k,n_n,T)&=&\frac{2}{3}A\left(1-x_0 \right)u\nonumber \\
&+&\frac{\frac{2}{3}B\left(1-x_3\right)u^{\sigma}}{(1+\frac{2}{3}B^{\prime}\left(1-x_3 \right)u^{\sigma-1})^2}\left[(\sigma+1) \right.\nonumber \\
&+&\left.\frac{4}{3}B^{\prime}\left(1-x_3\right)u^{\sigma-1}\right]\nonumber \\
&+&\sum_{i=1,2}\frac{2}{5n_0}\left[3C_i- 4Z_i \right]{\cal J}_i(n_n,T)\nonumber
\\
&+&\frac{2}{5}u\sum_{i=1,2}(3C_i - 4Z_i){\rm g}(k,\Lambda_i).
\label{eq4}
\end{eqnarray}


The primary source of momentum dependence in Brueckner theory arises from the nonlocal nature of the exchange interaction. As discussed by Bertsch et al.~\cite{Bertsch-1988}, a single-particle potential $U(n)$ that depends solely on baryon density is an oversimplification. Moreover, it is well established that nuclear interactions involve significant exchange effects, which introduce a momentum dependence to the single-particle potential, subsequently influencing the energy density functional. To conduct comprehensive studies of heavy ion collisions, Gale et al.~\cite{Gale-1987} proposed the following parametrization for the momentum component of the single-particle potential
\begin{equation}
 U(n,k) \sim C\frac{n}{n_0}\frac{1}{1+(k-\langle k'\rangle)^2/\Lambda^2}.  \label{eq9}
\end{equation}
The present model, which is a generalization of that
proposed by  Gale et al.~\cite{Gale-1987}, has been successfully applied
in heavy ion collisions and astrophysical studies over the
years (see Refs.~\cite{Prakash-1997,Li-2008} and references therein).

The first two terms of the right-hand side of Eq.~(\ref{eq4}) arise
from local contact nuclear interactions that led to power density
contributions such as in the standard Skyrme equation of state.
These are assumed to be temperature independent. Moreover,  the first term  represents an attractive interaction, while the second one  corresponds to a repulsive interaction that becomes dominant at high densities (for $n > 0.6$ fm$^{-3}$). 
The third term describes the
effects of finite-range interactions according to the chosen
function ${\rm g}(k,\Lambda_i)$, and is the temperature-dependent part of
the interaction. This interaction is attractive and important at
low momentum, but it weakens and disappears at very high
momentum.
By choosing the function (\ref{g-1}), we introduce two finite-range terms: one representing a long-range attraction and the other a short-range repulsion (for more details see also Refs.~\cite{Prakash-1997,Bertsch-1988,Gale-1987}). 
Moreover, in Eq.~(\ref{g-1}) we use consider that  $\Lambda_1 = 1.5k_{F{n_0}}$ and  $\Lambda_2 = 3k_{F{n_0}}$ (where  $k_{F{n_0}}$ is the neutron Fermi momentum at the saturation density $n_0$). 
The parameterization of  Eq.~(\ref{eq5}) is presented  in Table \ref{tab-1}.
The parameters $A$, $B$, $B'$ $\sigma$, $C_1$ and $C_2$, which
appear in the description of symmetric nuclear matter and the
additional parameters $x_0$, $x_3$, $Z_1$, and $Z_2$ used to determine
the properties of asymmetric nuclear matter, are treated as
parameters constrained by empirical knowledge (for more details see Ref.~\cite{Prakash-1997}). 
\begin{table}
\centering
\caption{The parametrization of the potential energy  $V_{\rm int}(n_n,T)$ given by Eq.~(\ref{eq5}) (for more details see Ref.~\cite{Prakash-1997}).}
\label{tab-1}    
\begin{tabular}{cccccccccc}
\hline
 $A$  & $B$ & $B'$& $\sigma$ &  $C_1$ & $C_2$ & $x_0$ &  $x_3 $ & $Z_1$ & $Z_2$  \\\hline
-46.65 & 39.45 & 0.3 & 1.663 & -83.84 & 23 & 1.654 & -1.112 & 3.81 & 13.16 \\\hline
\end{tabular}

\end{table}

Furthermore, the entropy density $s_n$ which is a basic ingredient of the equation of state, is given by  \begin{equation} s_n = -g_s \int  \frac{ d^3k}{(2\pi)^3}[f_n \ln f_n +(1 - f_n)(\ln(1-f_n))].
\label{eq12}\end{equation}
The pressure of hot neutron matter is now expressed by
\begin{equation} P_n(n_n,T)= \mu_n n_n + T s_n - {\cal E}_n(n_n,T).
\label{eq13}
 \end{equation}
In total, the energy density ${\cal E}_n(n_n,T)$
together with the pressure 
$P_n(n_n,T)$
fully characterizes the equation of state of hot  neutron matter.

\label{sec II}

\subsection{Hot quarkyonic matter }
\begin{figure}
\includegraphics[width=124pt,height=9.6pc]{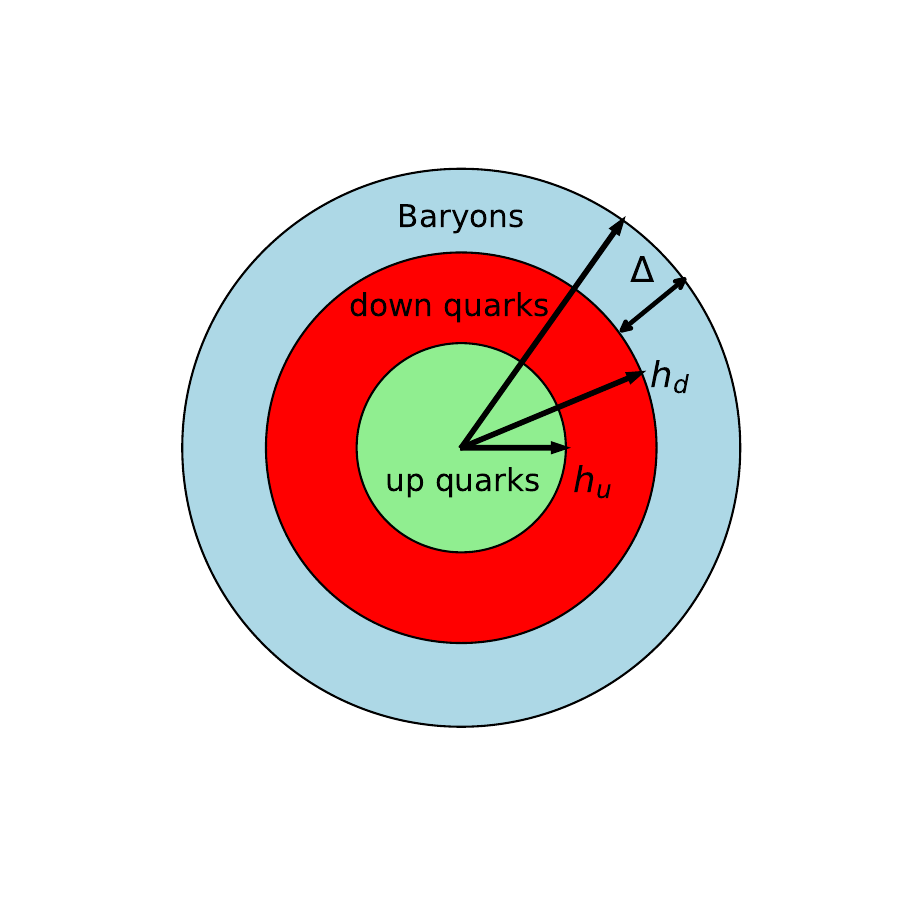} 
\includegraphics[width=118pt,height=9.4pc]{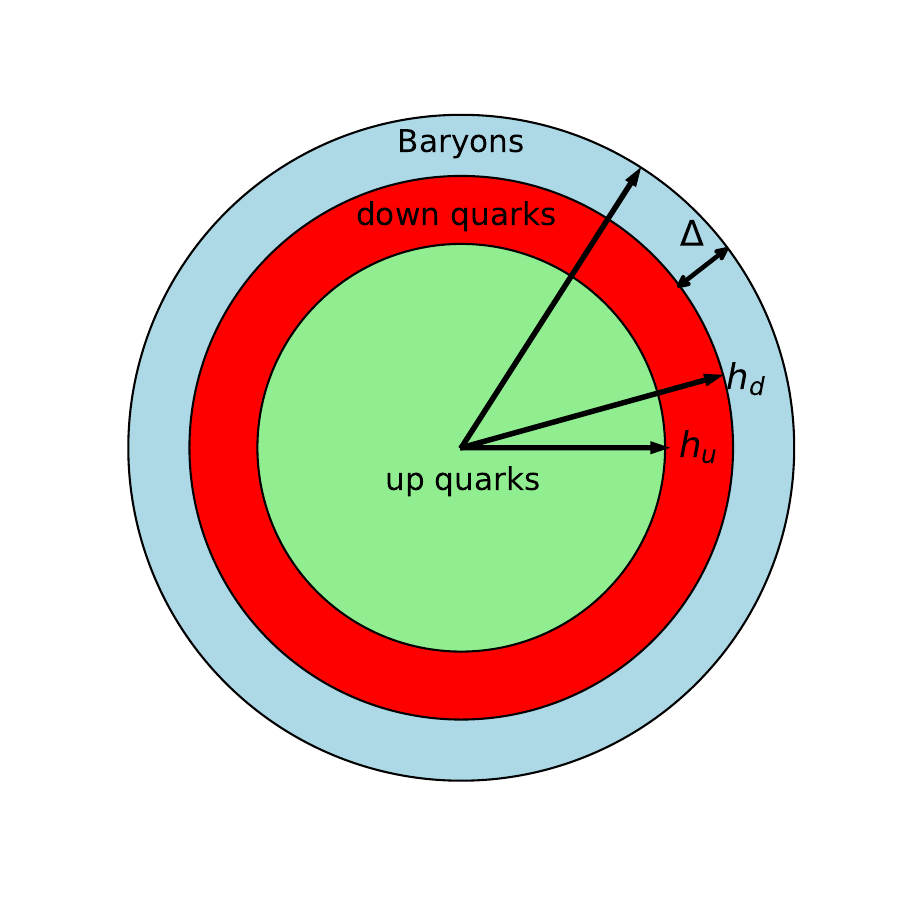}
\caption{The momentum space represenation  of quarkyonic matter at zero temperature (left) and at finite temperature (right)  respectively. Quarks occupy all states from zero momentum up to maximum momentum ($h_d$ and $h_u$) for the down and up
quark respectively). At higher momentum, quarks are confined into baryons that occupy momentum states in a shell with width $\Delta$ (see also the text and Ref.~\cite{PhysRevLett.122.122701} for more details). As the temperature increases, quarks spread out to higher momentum states and the width of the shell where quarks are confined into nucleons decreases.}
\label{Fermisea}
\end{figure}

Fig.~\ref{Fermisea} provides a schematic overview of the quarkyonic model.  In this representation, quarks occupy low-momentum states, whereas neutrons reside in high-momentum states. This distinction facilitates a clearer understanding of the respective contributions of quarks and neutrons to the total energy density.  
To investigate quarkyonic matter and its impact on neutron star properties, we begin with a simplified model where the neutron star is composed solely of neutrons, along with up and down quarks~\cite{PhysRevLett.122.122701}. This assumption simplifies the calculations and serves as a useful approximation, particularly considering that in a neutron star, the number of protons and electrons is minimal compared to the number of neutrons. As a result, neutrons and quarks play the primary role in determining the structure of a neutron star.

We assume that quarks are non-interacting, while neutrons interact exclusively with one another. Furthermore, we extend our calculations to quarkyonic matter at finite temperature, so that the system no longer behaves as a fully degenerate Fermi gas, as in the cold  case. Nevertheless, we maintain the requirement that low-momentum states are occupied by quarks, whereas higher-momentum states are occupied by neutrons, as illustrated schematically in Fig.~\ref{Fermisea}, consistent with the cold matter scenario. The primary distinction in the finite-temperature case is that the energy states of both nucleons and free quarks are distributed above the Fermi momentum, with the probability of occupying a given state governed by the Fermi–Dirac distribution. 

In our previous study, in the cold case ($T=0$)~\cite{Folias-2025}, the width of the momentum shell in which quarks are confined within nucleons was determined using the relation given in Ref.~\cite{PhysRevLett.122.122701}
\begin{equation}\Delta = \frac{\Lambda_{Qyc}^3}{\hbar^3 c^3 k_{F_n}^2 }+ \kappa \frac{\Lambda_{Qyc}}{\hbar c N_c^2}, \label{eq14-a}
\end{equation}
where we use the values   $\Lambda_{Qyc} \approx \Lambda_{QCD}$ and $\kappa =0.3$.
Moreover,  we assume that Eq. (\ref{eq14-a}) remains valid, in order to ensure that our model at low temperatures is consistent with the cold case and that the neutron fraction decreases at high densities, where quarks begin to appear. For higher temperatures, where nucleons spread out to higher momentum states, the Fermi momentum $k_{F_n}$ makes no sense anymore, so we replace it with an upper limit in the momentum space (we call it $h_n$) which can now have higher values than the Fermi momentum (see  Fig.~\ref{Fermisea}). One might expect that neutrons could occupy states up to arbitrarily high momenta. However, since the Fermi–Dirac distribution function becomes negligible at high momenta, we can safely assume this upper limit for neutron momentum states without compromising the validity of our calculations. Accordingly, we can rewrite Eq.~(\ref{eq14-a}) as follows
\begin{equation}\Delta = \frac{\Lambda_{Qyc}^3}{\hbar^3 c^3 h_n^2 }+ \kappa \frac{\Lambda_{Qyc}}{\hbar c N_c^2}. \label{eq14-b}
\end{equation}
Moreover, in order to ensure charge neutrality we consider that
\begin{equation}n_d = 2 n_u. 
\label{charge neutrality}
\end{equation}
Now, the quark number density and energy density are defined respectively, as follows 
\begin{equation}n_{i} = \frac{g_s N_c}{(2\pi)^3} 4\pi \int_0^{h_{i}} f_{i}  k^2 dk, \quad i=u,d
\label{eq15}\end{equation}
\begin{equation}{\cal E}_{Q}(n_i,T) = \frac{g_s N_c}{(2\pi)^3}\sum_{i=u,d}4\pi \int_0^{h_{i}}   f_{i}\sqrt{(\hbar c k)^2+ m_{i}^2c^4} k^2 dk,
\label{eq16}
\end{equation}
where $g_s$ is the degeneracy factor for the spin and $N_c$ is the number of colors for quarks. The Fermi-Dirac distribution function of quarks  is  given by 
\begin{equation}f_{i} = \frac{1}{e^{(E_{i} - \mu_{i})/T }}.
\label{fQ} \end{equation}
$E_i$ stands for the single particle energy for up and down quarks
\begin{equation}E_{i} = \sqrt{(\hbar c k)^2+ m_{i}^2c^4} \label{fQ} 
 \end{equation}
where  $\mu_i$  represents the associated chemical potential.

The number density $n_n$ of  neutrons, in quarkyonic matter,  will be given by the the expression   
\begin{equation}n_n =\frac{g_s} {(2\pi)^3} 4\pi \int_{h_d}^{h_d+ \Delta}  f_n k^2 dk\label{eq17}\end{equation}
and the corresponding energy density by  
\begin{eqnarray}{\cal E}_{n}(n_n,T) &=&\frac{g_s}{(2\pi)^3}4\pi \int_{h_d}^{h_d + \Delta} f_n \sqrt{(\hbar c k)^2+ m_{n}^2c^4}k^2 dk \nonumber\\
&+& V_{\rm int}(n_n,T). \label{eq18}
\end{eqnarray}
%
The term $V_{\rm int}(n_n,T)$ in Eq. (\ref{eq18}), is given by Eq. (\ref{eq5}) where now  the integrals ${\cal J}_i(n_n,T)$ are defined as
\begin{eqnarray}
{\cal J}_i(n_n,T)&=&\frac{2}{(2\pi)^3}\int d^3k f_n {\rm g}(k,\Lambda_i)\\
&=&\frac{2}{(2\pi)^3} 4\pi\int_{h_d}^{h_d + \Delta}  f_n \left[1+\left(\frac{k}{\Lambda_i}  \right)^2 \right]^{-1}   k^2 dk. 
\nonumber\label{εq19}
\end{eqnarray}
The corresponding  entropy densities  for each quark flavor (solely up and down quarks), are given by 
\begin{equation} s_{i} =  -\frac{g_s N_c}{(2\pi)^3} 4\pi \int_0^{h_{i}}  [f_{i} \ln f_{i} +(1 - f_{i})(\ln(1-f_{i}))]k^2 dk
\label{eq23}
\end{equation}
 while this for neutrons from the
\begin{equation} s_n =  -\frac{g_s}{(2\pi)^3} 4\pi\int_{h_d}^{h_d + \Delta}  [f_n \ln f_n +(1 - f_n)(\ln(1-f_n))]k^2 dk.
\label{eq24}
\end{equation}

The timescales of weak and strong interactions are much shorter than the timescales of the evolution of a neutron star, so a reasonable consideration is to we require the chemical equilibrium relation for neutrons and quarks chemical potentials
\begin{equation}
\mu_n = \mu_u + 2 \mu_d. 
\label{eq25}
\end{equation}

Quark masses are obtained by $m_i = m_N/N_c$ ($ i=u,d$) where $m_N$ is the nucleon mass and $N_c$ is the number of colors. We set the number of colors and the degeneracy of the spin equal to $N_c = 3 $ and $g_s = 2$ respectively and we introduce the quarkyonic matter at a baryon density around $n_B = 0.2 - 0.3$ fm$^{-3}$. The total baryon density and total energy density will be, respectively
\begin{equation} n_B = n_n + \frac{(n_u + n_d)}{3} \label{eq26}\end{equation}
 and
\begin{equation} {\cal E}_{\rm tot}(n_n,n_i,T)= {\cal E}_n(n_n,T) + {\cal E}_Q(n_i,T). 
\label{eq27}
\end{equation}
Finally, the total pressure is  given   by
\begin{equation} 
P_{\rm tot}(n_n,T) = \sum_{i = n,u,d}\left(\mu_i n_i + T s_i\right) - {\cal E}_{\rm tot}(n_n,n_i,T). \label{eq28}
\end{equation}
$ {\cal E}_{\rm tot}$ and $P_{\rm tot}$ define  the equation of state of hot quarkyonik matter. This quantity constitutes the fundamental ingredient for investigating the properties of both hot quarkyonic matter and neutron stars composed of this type of matter.

{\it  Calculation recipe}:
We assume that quarks occypy momentum states from zero up to a maximum momentum $h$ ($h_u$ for up quarks and $h_d$ for down) and neutrons occupy a Fermi shell in the momentum space from h up to a maximum momentum not fixed, as in the cold case, due to the effect of the temperature 
(see Fig.~\ref{Fermisea}). We start our calculations from Eq. (\ref{eq15}): For a fixed value of T and  for a given value of up quark number density $n_u$ we solve iteratively Eq. (\ref{eq15}) to compute for several values of the upper momentum $h_u$ the up quark chemical potential $\mu_u$. Once the solutions for the chemical potentials converge to stable values, we determine the up-quark chemical potential and the corresponding maximum momentum state $h_u$ associated with that density. Then we impose the charge neutrality, that is  Eq.~(\ref{charge neutrality}),  so to compute the down quark number density $n_d$. We repeat the previous procedure, to compute iteratively the chemical potential for down quarks ($\mu_d$). After that we compute from chemical equilibrium relation the chemical potential for neutrons. Then, we solve iteratively Eq.~(\ref{eq17}) to compute the neutrons number density. So,  the total baryon density is given from Eq.~(\ref{eq26}). Since the chemical potentials and number densities are all known, we can easily compute the energy densities, entropy densities for neutrons and quarks respectively and finally we calculate the total pressure from Eq.~(\ref{eq28}). We repeat this process for a wide range of up quark number densities (so for several values of the total baryon density) and so we construct equations of state for a specific temperature. Finally we repeat our calculations for seceral values of the temperature both for pure neutron and quarkyonic matter.

\section{Neutron star structure}
The bulk properties of neutron stars, such as mass, radius and tidal deformability are determined by solving the coupled Tolman–Oppenheimer–Volkoff (TOV) equations~\cite{Shapiro:1983du,Haensel2007NeutronS1,schaffner-bielich_2020}.  As discussed in the preceding section, the equation of state serving as the fundamental input to the TOV equations, is constructed using the proposed quarkyonic matter model, which encompasses both the cold and finite-temperature regimes.

In recent years, valuable insights have been gained from observing gravitational waves produced by the mergers of black hole–neutron star and neutron star–neutron star binary systems. These events provide an opportunity to measure various properties of neutron stars. Notably, during the inspiral phase of binary neutron star systems, tidal effects can be detected. More specifically, the tidal Love number $k_2$ 
characterizes how a neutron star responds to an external tidal field, depending on both its mass and the applied equation of state. The exact relation governing these tidal effects is presented below~\cite{Flanagan-2008,Hinderer-2008}
\begin{equation}
Q_{ij}=-\frac{2}{3}k_2\frac{R^5}{G}E_{ij}\equiv- \lambda E_{ij},
\label{Love-1}
\end{equation}
where $\lambda$ is the tidal deformability and $E_{ij}$  the applied tidal field. 
In addition, an important and well measured quantity by the gravitational wave detectors, which can be treated as a tool to impose constraints on the EoS, is the dimensionless tidal deformability $\Lambda$, defined as 
\begin{equation}
    \Lambda=\frac{2}{3}k_2 \left(\frac{c^2R}{GM}\right)^5=\frac{2}{3}k_2 (1.473)^{-5}\left( \frac{R}{{\rm Km}} \right)^5\left(\frac{M_{\odot}}{M}  \right)^5
\end{equation}
We notice that $\Lambda$ is sensitive to the neutron star radius, hence can provide information for the low-density part of the EoS, which is also related to the structure and properties of finite nuclei.

\section{Results and Discussion}

\begin{figure}
\includegraphics[width=250pt,height=18pc]{
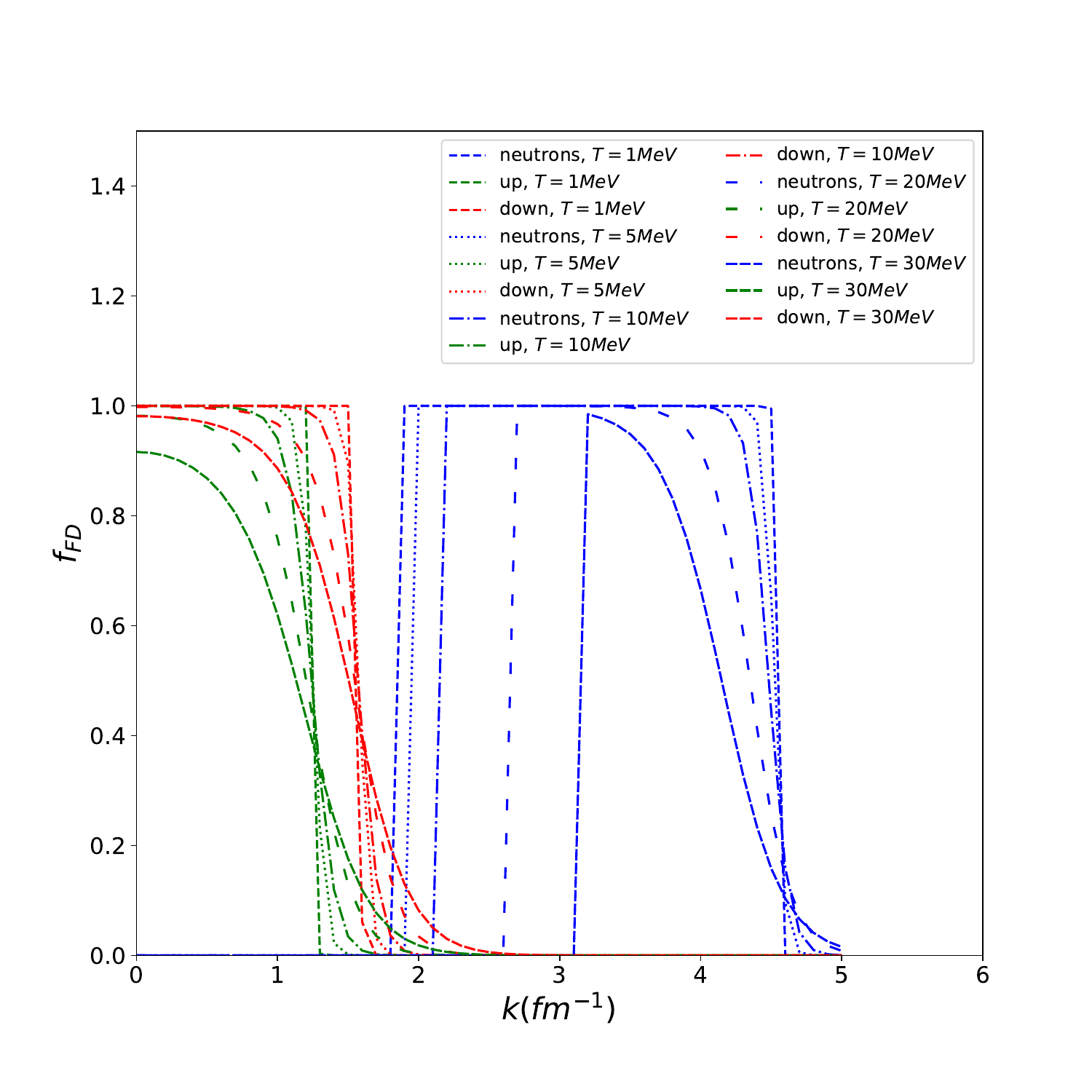}
\caption{The Fermi - Dirac distribution functions for quarks and neutrons, in quarkyonic matter, at total baryon density $n_B = 0.3$ fm$^{-3}$ and for several values of the temperature.}
\label{F-D}
\end{figure}
\begin{figure}
\includegraphics[width=250pt,height=18pc]{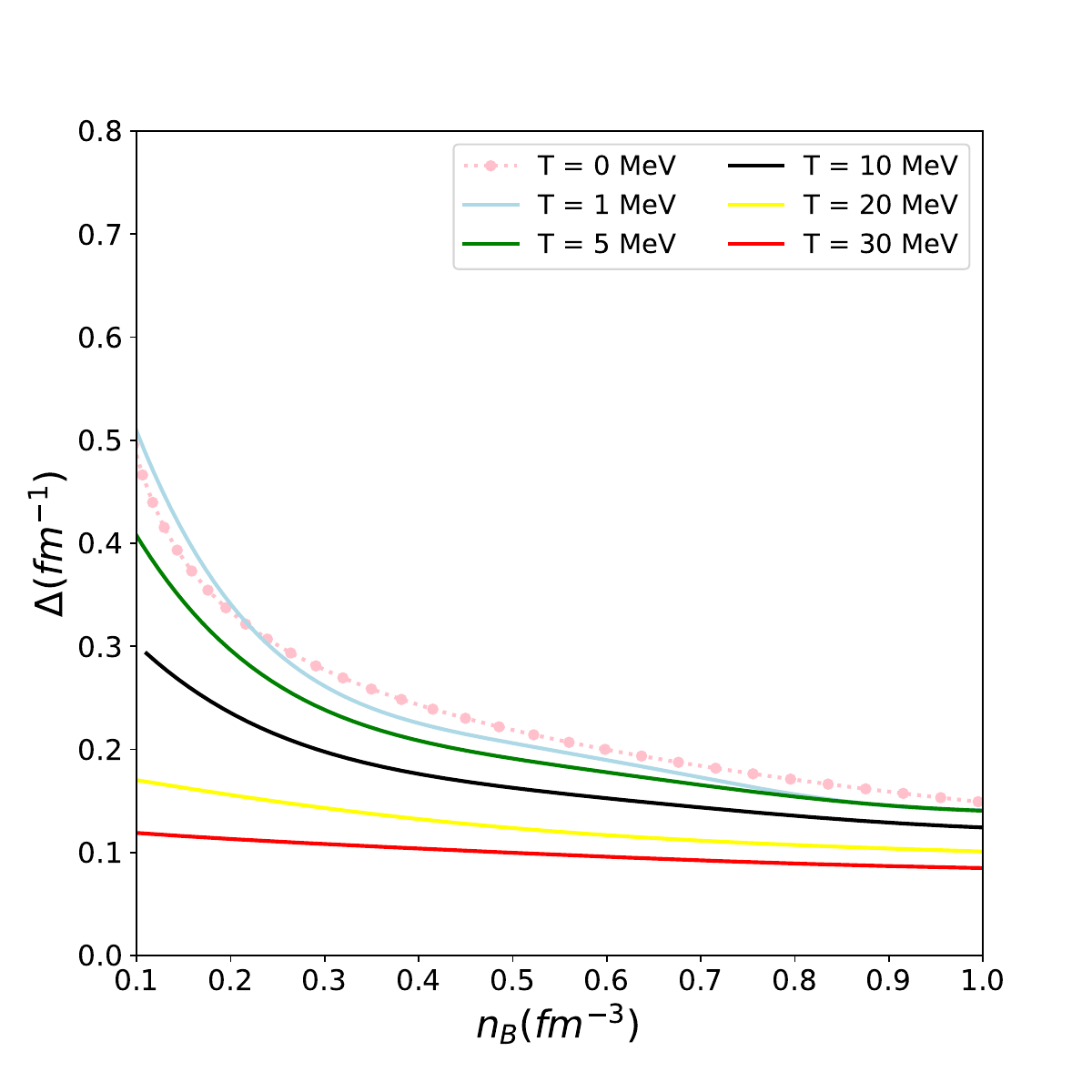}
\caption{The width of the momentum shell $\Delta$ as a function of the total baryon density $n_B$, for several values of the temperature, for $\kappa =0.3 $, $N_c = 3$ and $\Lambda_{Qyc} = 200$ MeV .}
\label{Shell}
\end{figure}

\begin{figure}
\includegraphics[width=250pt,height=18pc]{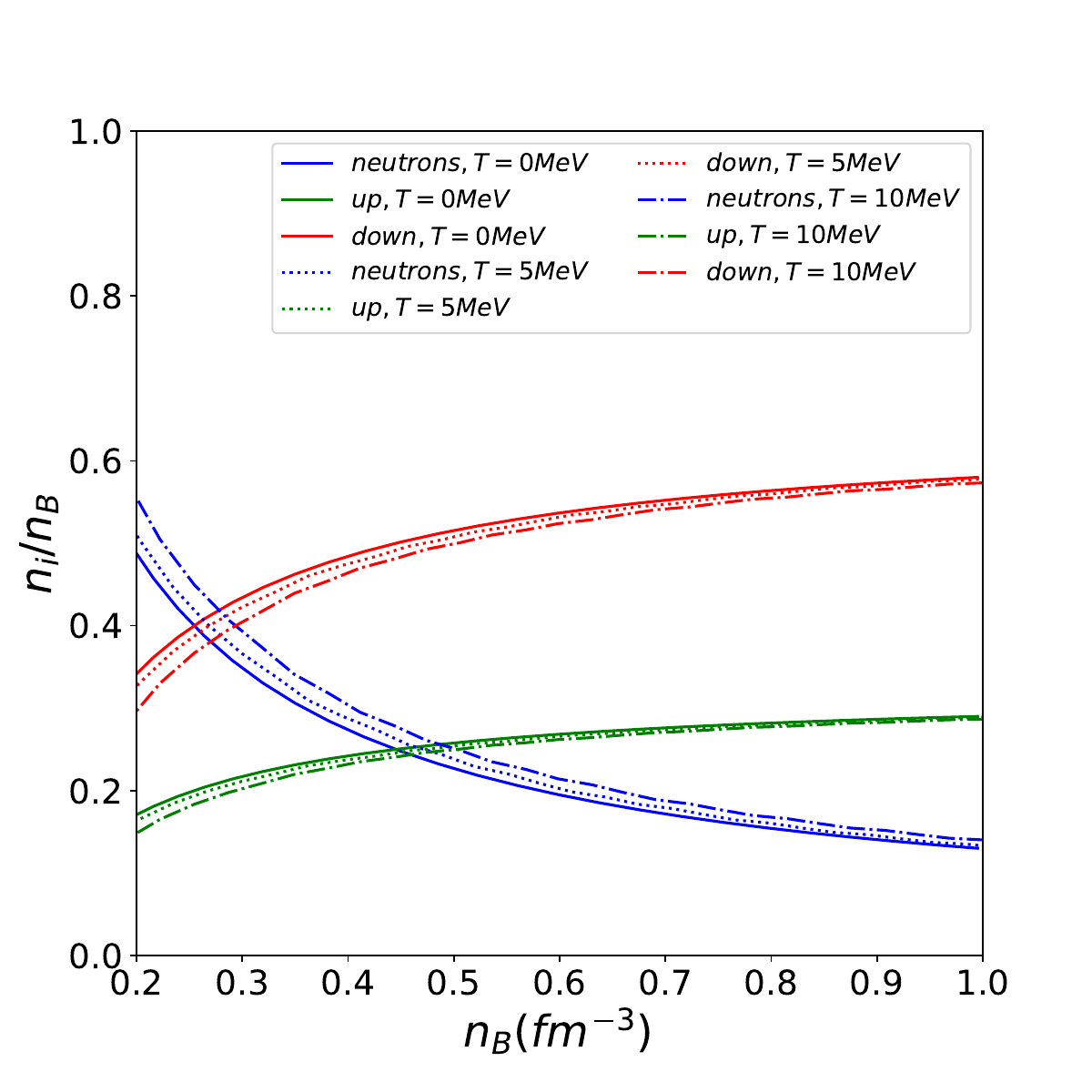}
\caption{Neutron, up and down quark number densities as a fraction of the total baryon density, for cold quarkyonic matter and for finite temperature quarkyonic matter for $T = 5, 10 \ {\rm MeV}$.}
\label{particle fractions}
\end{figure}

\begin{figure}
\includegraphics[width=250pt,height=18pc]{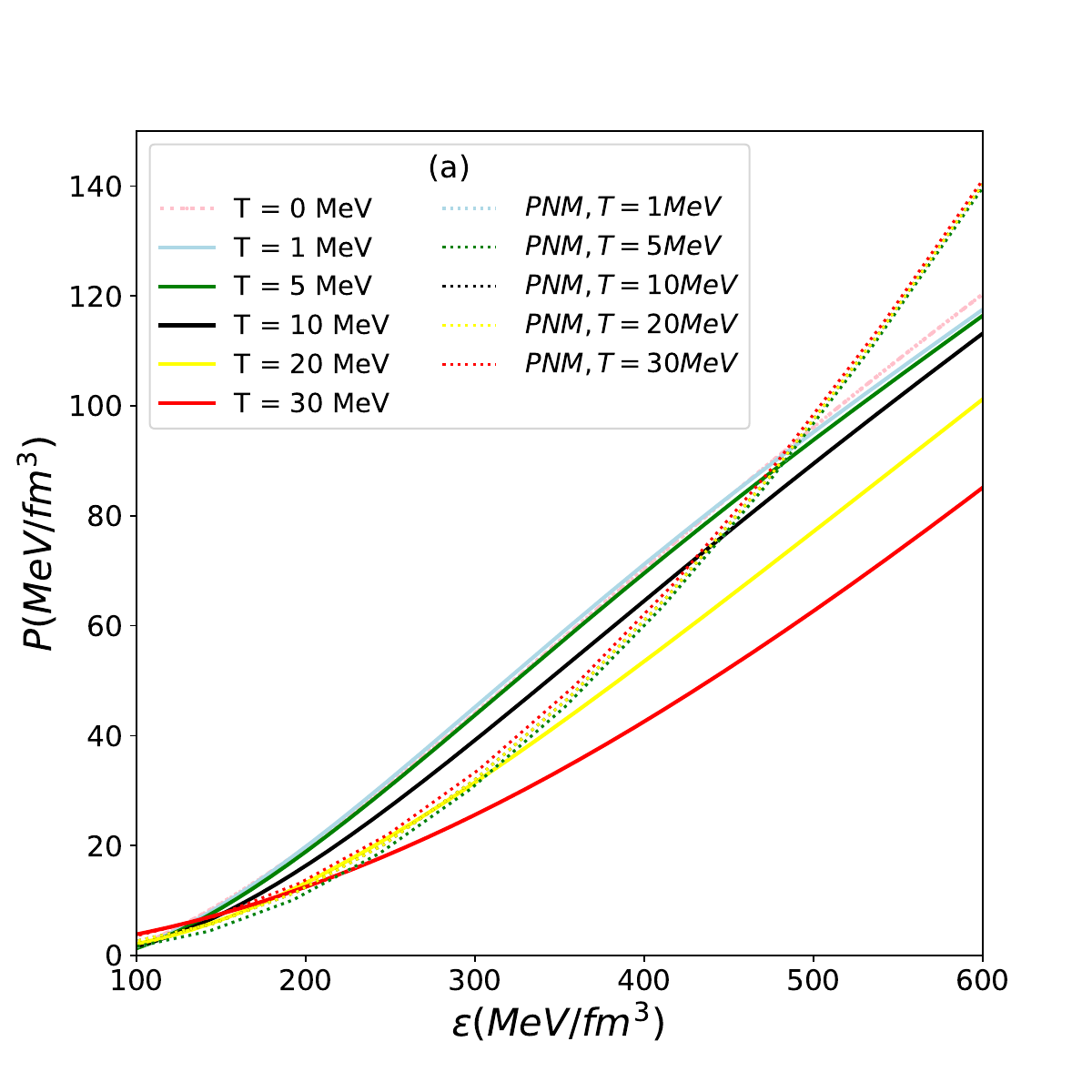}
\includegraphics[width=250pt,height=18pc]{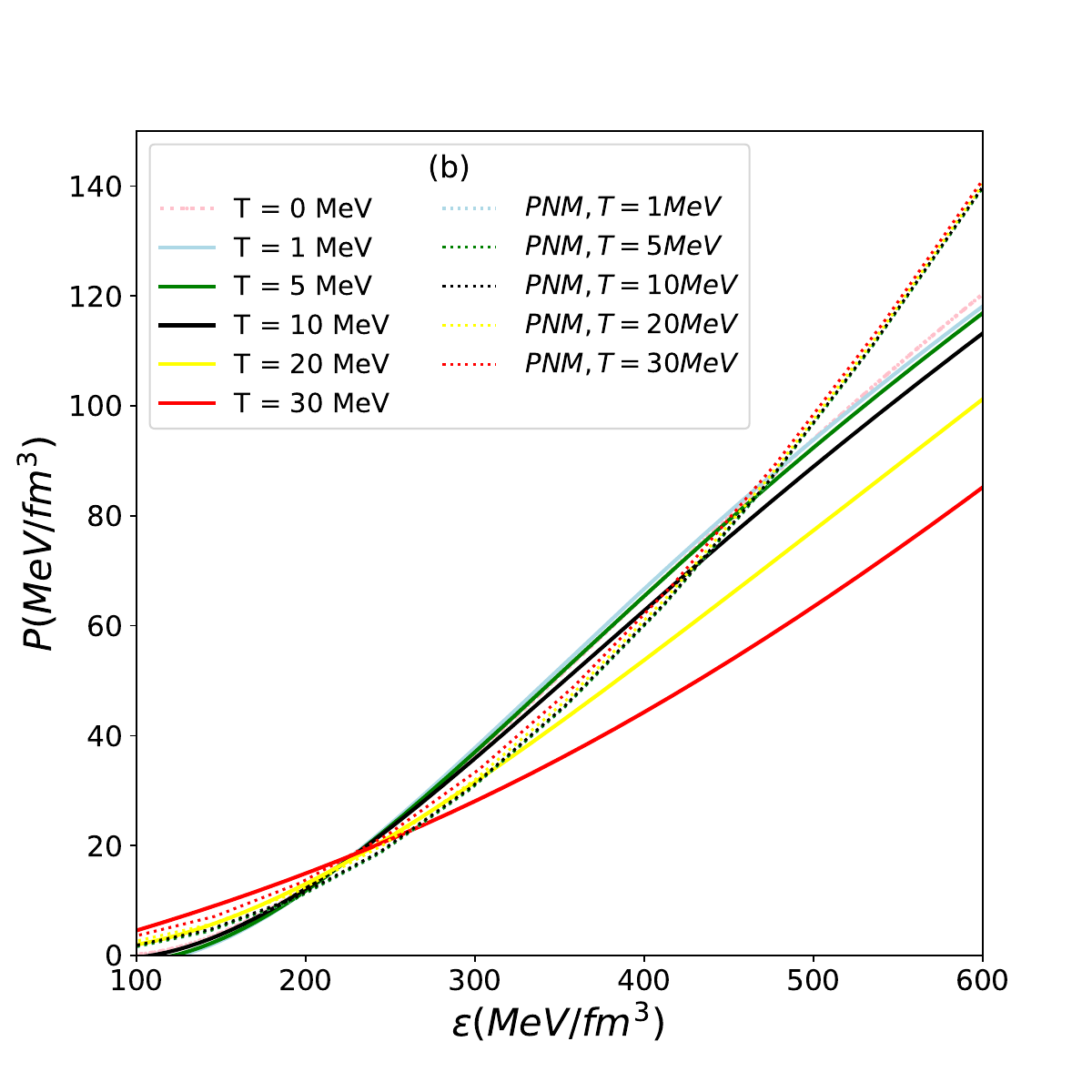}
\caption{Equations of state for pure neutron matter and quarkyonic matter with:   (a) $n_{\rm tr}=0.2\ {\rm fm}^{-3}$  and (b)  $0.3 \ {\rm fm}^{-3}$, for several values of temperature.}
\label{EOS}
\end{figure}

\begin{figure}
\includegraphics[width=250pt,height=18pc]{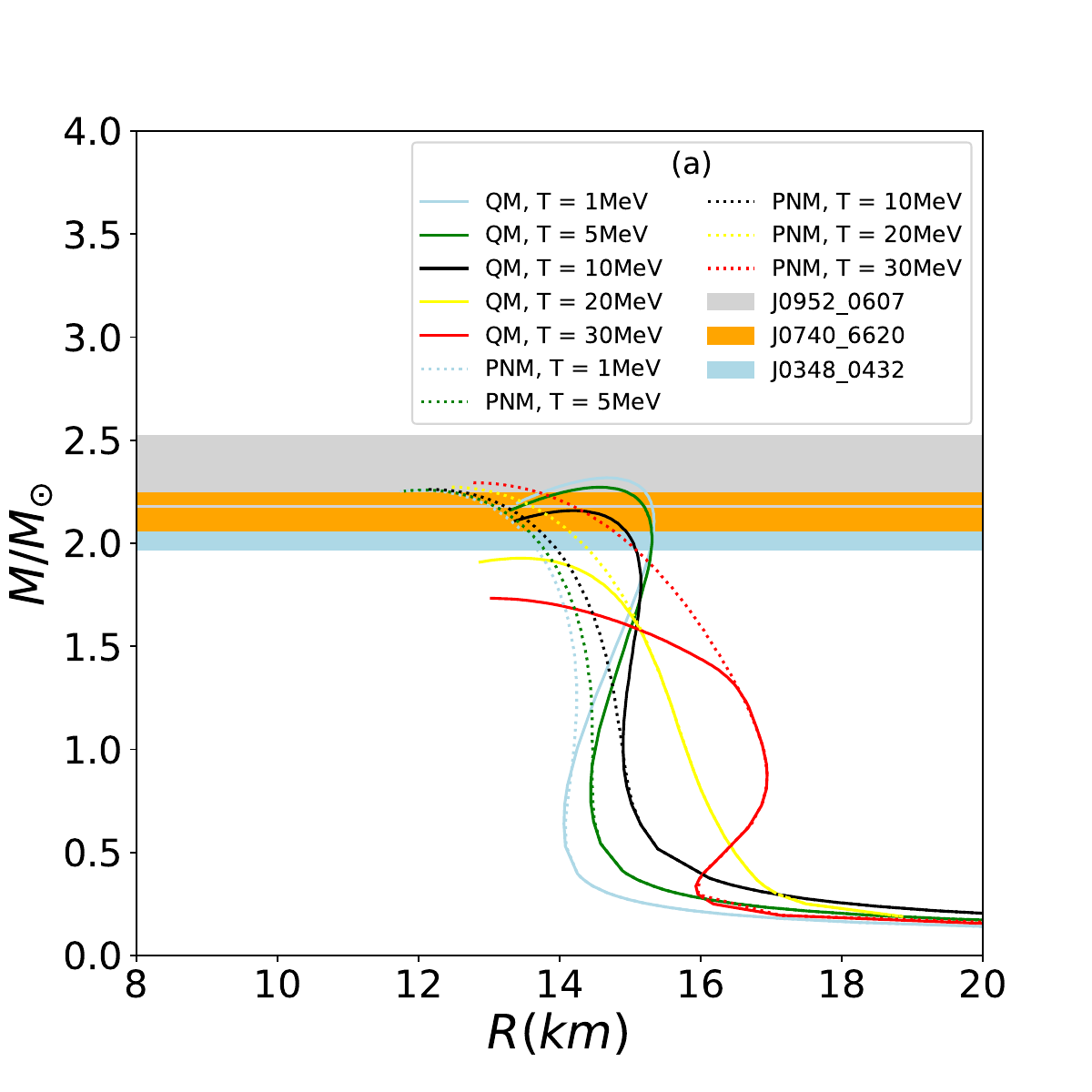}
\includegraphics[width=250pt,height=18pc]{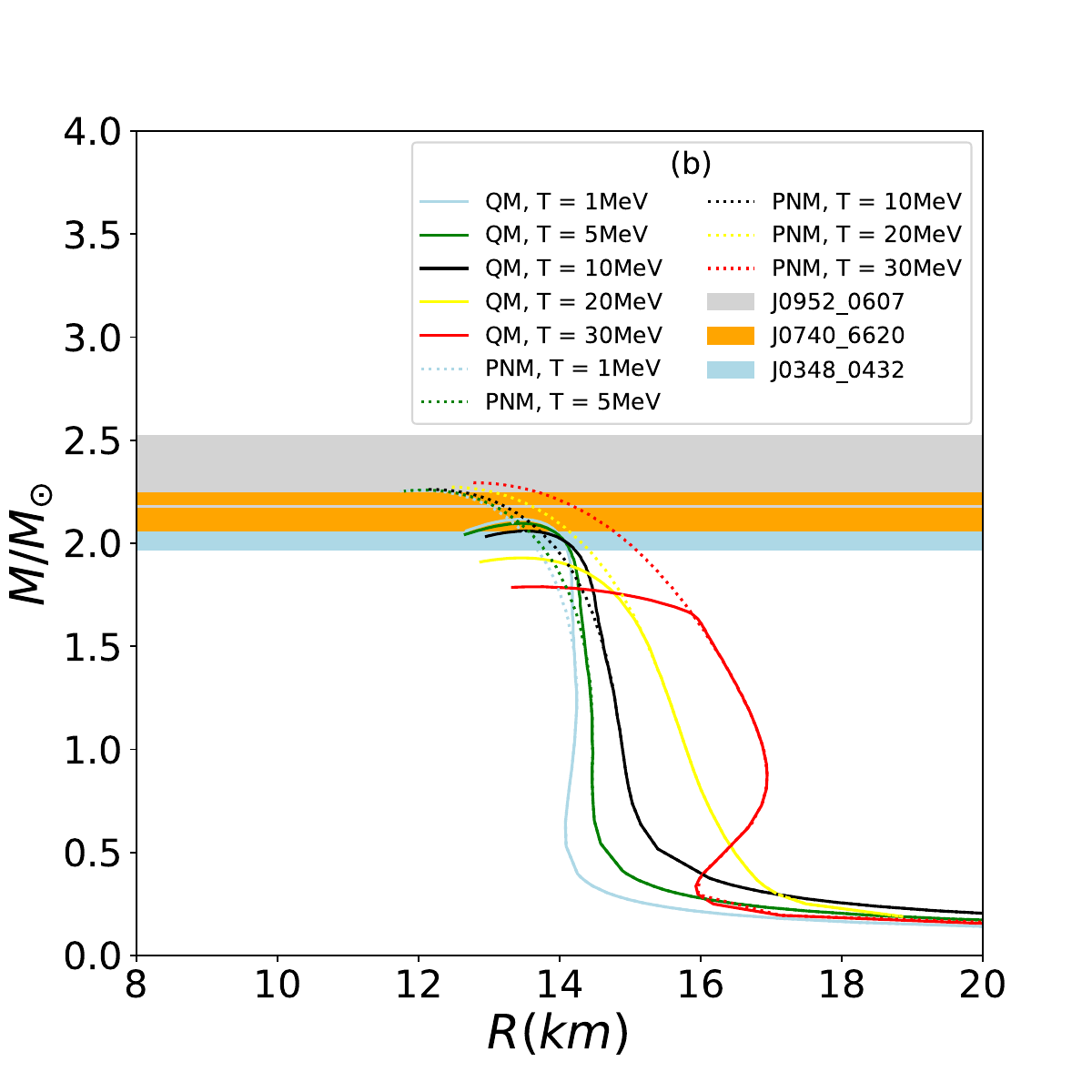}
\caption{M-R diagrams for pure neutron matter and  quarkyonic matter with: (a) $n_{\rm tr}=0.2\ {\rm  fm^{-3}}$ and (b) $0.3\  {\rm  fm^{-3}}$, for several values of the temperature. The shaded regions correspond to possible constraints on the maximum mass from the pulsar observation of PSR J0348+0432, PSR J0740+6620 and PSR J0952+0607~\cite{7712d035a7c043dd8e3205ecb6703f51,Antoniadis:2013pzd,PhysRevD.109.063017,2022ApJ...941..150S,Miller:2021qha}.}  
\label{mass-radius}
\end{figure}

\begin{figure}

\includegraphics[width=250pt,height=18pc]{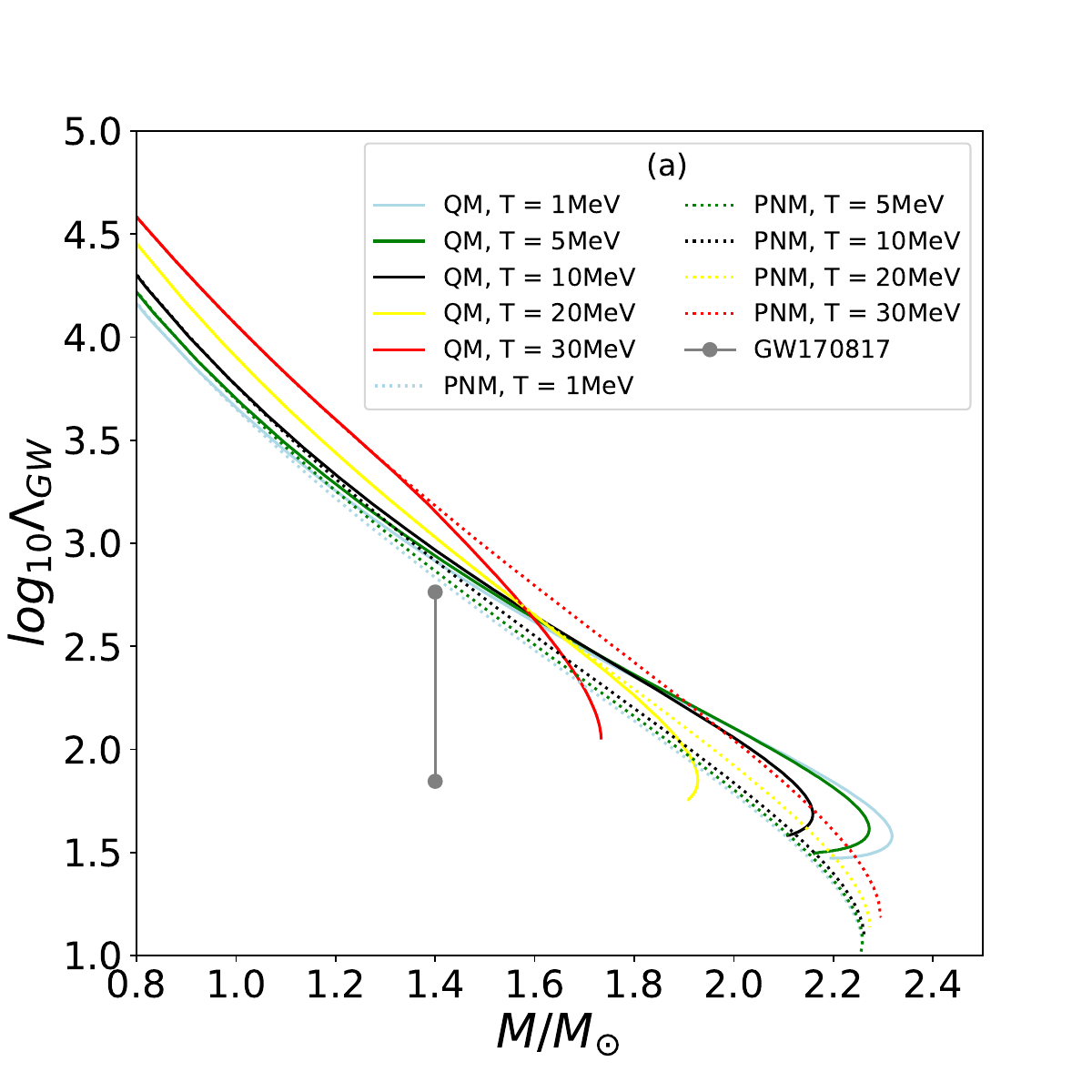}
\includegraphics[width=250pt,height=18pc]{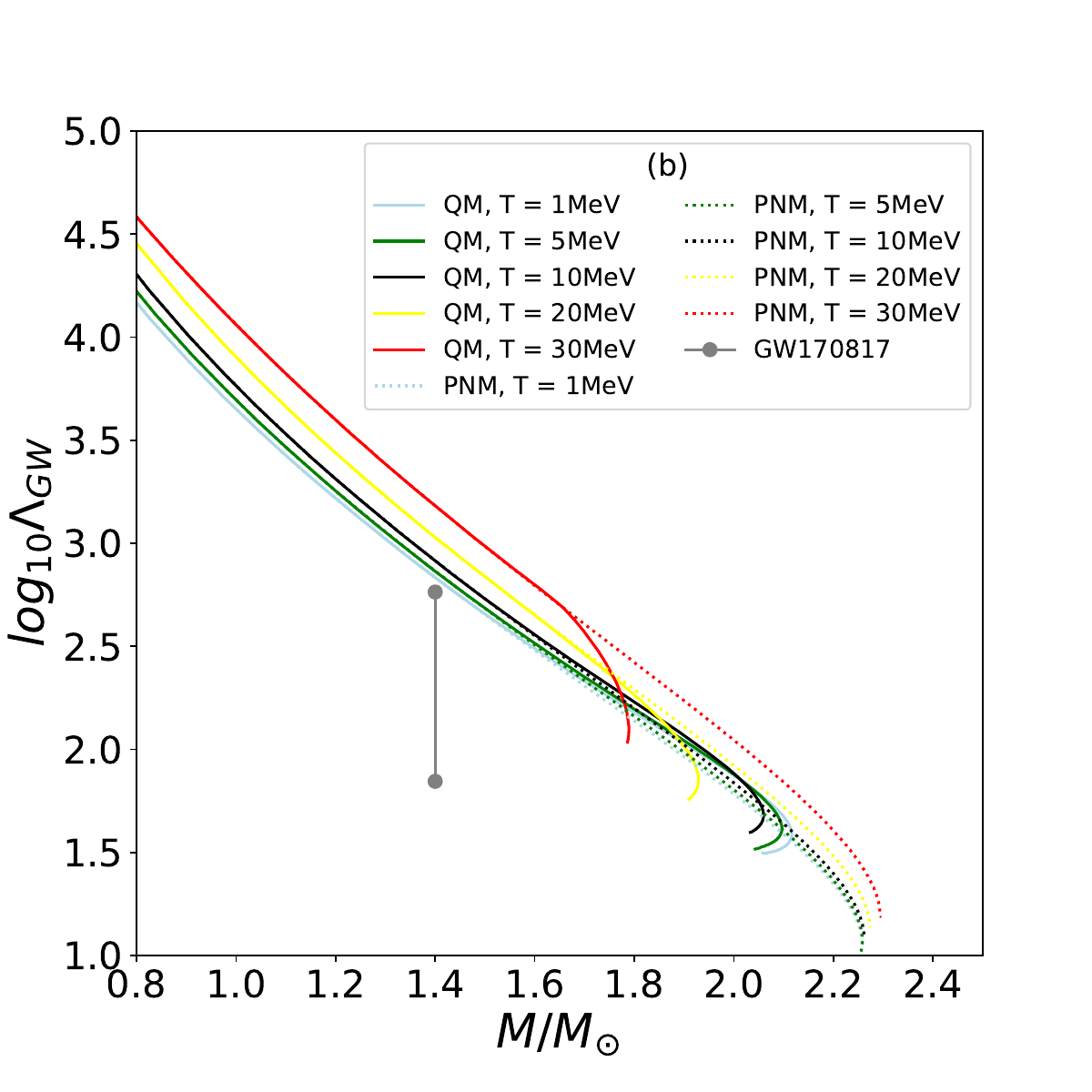}
\caption{The logarithm of the tidal deformability $\Lambda$ for pure neutron matter and  quarkyonic matter with: (a) $n_{\rm tr}=0.2\ {\rm fm^{-3}}$ and (b)  $0.3\  {\rm  fm^{-3}}$, for several values of the temperature. The grey line corresponds to the event GW170817~\cite{Abbott_2017}.}
\label{tidal}
\end{figure}

Investigating the influence of temperature on the fundamental properties of neutron stars, whether pertaining to static or dynamic phenomena, is not only necessary but also crucial for achieving a realistic description of these objects. In particular, the effect of temperature on quarkyonic matter is an issue that has not been addressed in the relevant literature (with the exception Ref.~\cite{Sen-2021}). In the present study, we attempt to examine the influence of temperature on the properties of hot quarkyonic matter by employing a methodology distinct from that used in Ref.~\cite{Sen-2021}. Furthermore, extending the framework established in that reference, the present analysis incorporates the effects of thermal quarkyonic matter on the fundamental properties of neutron stars.

More specifically, an attempt was made to introduce temperature in a self-consistent manner (such that, in the limit as the temperature approaches zero, the results converge to those of cold quarkyonic matter). In addition, it was assumed that the increase in temperature, which drives the free quark gas to higher momenta, does not overlap with the occupation of these states by neutrons. Consequently, this naturally leads to neutrons beginning to occupy momentum states starting from higher minimum values compared to the cold case. This rearrangement of the distributions in momentum space is not arbitrary, but rather emerges from the effect of temperature, while in all cases respecting chemical equilibrium, which is expressed through the equations of the chemical potentials of the participating particles 
(see Eq.~(\ref{eq25})).

In Fig.~\ref{F-D} we plot the Fermi–Dirac distribution functions, for quarks and neutrons in quarkyonic matter, at several temperatures. The occupation probability is shown as a function of momentum. At low temperature, the distributions exhibit sharp Fermi surfaces, while increasing temperature leads to a smoothening of the step function. Differences between quark and neutron distributions reflect their distinct Fermi momenta and chemical potentials in quarkyonic matter. The most salient feature is that increasing temperature drives the quarks to higher momenta, while simultaneously the neutrons begin to occupy higher-momentum states within quarkyonic matter, thereby avoiding potential overlap between the two. This behavior evidently has consequences for the contribution of the two aforementioned components to the equation of state of quarkyonic matter. 

To be more specific, in Fig.~\ref{Shell} we plot the width of the momentum shell as a function of the total baryon density, for several values of the temperature. In this case, the reduction of the momentum shell with increasing temperature is evident, particularly at low densities. As a consequence, the contribution of neutrons is significantly suppressed at low densities and high temperatures. At higher densities, this effect becomes smoother, although the influence of temperature remains non-negligible. In Fig.~\ref{particle fractions}, we plot the neutron as well as the up- and down-quark number densities, expressed as fractions of the total baryon density, for both cold quarkyonic matter and quarkyonic matter at finite temperature 
($T=5$ and $10$ MeV). It is evident that temperature effects are significant only at low baryon densities and become progressively less pronounced at higher densities. The dominance of up quarks is already apparent at low densities, while at higher densities the contribution of down quarks also becomes appreciable in comparison with the neutron densities.

In Fig.~\ref{EOS} we present the equations of state of quarkyonic matter, obtained within the present model, in comparison with those of pure neutron matter, for various temperatures. We consider two cases corresponding to transition densities $n_{\rm tr}=0.2\ {\rm  fm^{-3}}$ and $0.3\  {\rm fm^{-3}}$.  It is expected that the choice of the transition density will also affect the fundamental properties of the star, such as its mass and radius.
A particularly interesting finding is that, while the effect of temperature on pure neutron matter is observable but not significant (both at low and high densities), in the case of quarkyonic matter the impact is dramatic, especially at high densities, though still present at low densities. 

The effects of temperature, on the bulk hot neutron stars properties, are reflected in the corresponding M–R diagrams, as shown in Fig.~\ref{mass-radius}. We observe that in the case of pure neutron matter an increase in temperature leads to a stiffer equation of state (and consequently to larger maximum masses and corresponding radii).
In the case of quarkyonic matter, the results exhibit sensitivity not only to temperature but also to the chosen values of the transition densities. Specifically, lower transition densities lead to a stiffer quarkyonic equation of state, resulting in higher corresponding values of the maximum mass and radius. Conversely, increasing the temperature yields a softer equation of state, thereby reducing the corresponding maximum mass and radius.
It is also noteworthy, albeit expected, that the lower the transition density, the greater the deviation between the predictions of the equation of state for pure neutron matter and that of quarkyonic matter. For instance, at a transition density of $n_{\rm tr}=0.2\ {\rm  fm^{-3}}$, the predictions of the quarkyonic matter model differ significantly for stellar masses around 1.4 solar masses, which represent the majority of observed neutron stars. Conversely, when the transition occurs at higher densities, no noticeable deviation is observed in this mass range. It is also worth noting here that the reliability of the model at high temperatures remains uncertain. A more elaborated and refined approach could yield more credible and physically realistic results.

As observed from the M–R diagrams, particularly in the case of transition density  $n_{\rm tr}=0.2\ {\rm  fm^{-3}}$, the quarkyonic phase leads to larger stellar radii. This behavior is expected to be reflected in the values of the tidal deformability, which is highly sensitive (scaling with the fifth power) to the stellar radius. 
Thus, in Fig.~\ref{tidal},  we display the dependence of the tidal deformability on mass for various temperatures, for both pure neutron matter and quarkyonic matter. Additionally, we include constraints on the values of $\Lambda$ corresponding to a 1.4-solar-mass neutron star, as obtained from the GW170817 event~\cite{Abbott_2017}. It is evident that increasing the temperature leads to a significant expansion of the stellar radius (for a given mass) and, consequently, to a corresponding increase in the tidal deformability. Moreover, a reduction in the transition density enhances this behavior, extending it to even lower masses. Overall, the simultaneous increase in temperature and decrease in transition density play a decisive role in raising the tidal deformability over a wide range of stellar masses.

\section{Concluding Remarks}
The main conclusions of the present  work can be summarized as follows: (a) The study of temperature effects in quarkyonic matter is extremely limited, even though several of its important applications, such as in heavy-ion collisions, in the merger process of binary neutron-star systems, and in their cooling evolution, require thermal equations of state, (b) The impact of temperature on the structure of quarkyonic matter, at least within the framework of the present approach, is intricate and complex. The diffusion of quarks and neutrons to higher momenta alters the structure of the corresponding equation of state, both at low and high densities, (c) In particular, although the  temperature does not have a dramatic effect on the number densities of the various particles (particularly at high densities), it does significantly influence their distribution in momentum space and consequently, their contributions to the energy and pressure,
(d) The effect of temperature on the equations of state results in deviations from the cold case, leading to progressively decreasing maximum masses and, at the same time, increasing corresponding radii, (e) In any case, the transition density from the hadronic to the quarkyonic phase exerts a profound influence on the equation of state and, consequently, on the macroscopic properties of neutron stars. The interplay between temperature and transition density may give rise to intriguing stellar configurations, distinct from those corresponding to the purely hadronic phase,
(f) It is possible that the present model may not be sufficiently reliable at very high temperatures, as it leads to an almost complete suppression of the neutron contribution, an outcome that does not appear to have a physical basis, but rather seems to be an artificial feature of the approach. In this case, one possible approach would be to allow for an overlap between the quark and neutron distributions in momentum space. Such a treatment would, of course, require the introduction of additional assumptions and approximations.

\end{document}